\documentclass{article}
\usepackage{xcolor}
\usepackage{authblk}
\usepackage{wrapfig}
\usepackage{calrsfs}
\usepackage{caption}
\usepackage{graphicx}
\usepackage{epsf}
\usepackage{a4}

\begin{document}

\title{The Science of Murray Gell-Mann}
\author[a,b]{Sanjay Jain}
\author[c]{Spenta R. Wadia}
\affil[a]{Department of Physics and Astrophysics\\
University of Delhi, Delhi 110007, India}
\affil[b]{Santa Fe Institute, 1399 Hyde Park Road, Santa Fe, NM 87501 USA}
\affil[c]{International Centre for Theoretical Sciences\\
Tata Institute of Fundamental Research, Bangalore 560089, India}

\date{August 2019}
\maketitle

\centerline{\bf Abstract}
\begin{center}
\parbox{15truecm}{This article summarizes some of the most important scientific contributions of Murray Gell-Mann (1929-2019). (Invited article for {\it Current Science}, Indian Academy of Sciences.)}
\end{center}

\section{Murray Gell-Mann}
\begin{wrapfigure}{l}{0.25\textwidth}
\includegraphics[width=1.0\linewidth]{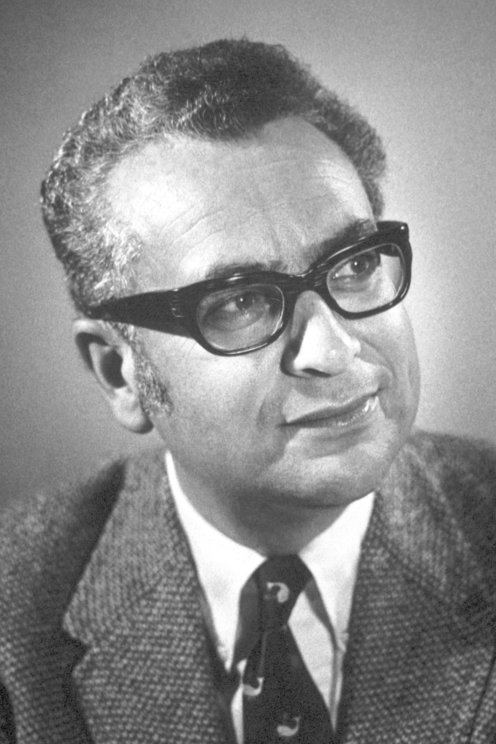} 
\caption*{\it Murray Gell Mann\\ (Photo credit: Nobel Foundation)}
\end{wrapfigure}

Murray Gell-Mann was among the very eminent and influential physicists of the second half of the 20th century. He was born in New York City on 15 September 1929. He graduated from Columbia Grammar School at the age of 14 and got his undergraduate degree from Yale University at 18. Gell-Mann obtained his doctorate degree from MIT, under the supervision of Victor Weisskopf, in just two and a half years. He joined the Institute for Advanced Study, Princeton in 1951 where he worked with Francis Low. During 1952-54 he was on the faculty of the University of Chicago with Enrico Fermi and Marvin Goldberger. He joined Caltech in 1955 as an associate professor on the recommendation of Richard Feynman, where he continued until his retirement in 1993 as Robert Andrews Millikan Professor of Theoretical Physics. 1993 onwards he remained the R.A. Millikan Professor Emeritus at Caltech and Distinguished Fellow at the Santa Fe Institute. He was awarded the Nobel Prize in Physics in 1969 ``for his contributions and discoveries concerning the classification of elementary particles and their interactions". His students include Sydney Coleman, James Hartle, Kenneth G. Wilson, Christopher T. Hill and Barton Zweibach.  He passed away in Santa Fe on 24 May 2019.

Gell-Mann made pathbreaking contributions to elementary particle physics. He produced a framework to describe high energy particles and their interactions by proposing a new quantum number `strangeness' and an organizing principle he called `eight-fold way'. He made the proposal of `quarks' as elementary building blocks of all known matter except particles like electron and neutrino.  On the occasion of Gell-Mann's Nobel Prize in 1969, Richard Feynman said, ``Our knowledge of fundamental physics contains not one fruitful idea that does not carry the name of Murray Gell-Mann". Among his other seminal contributions was the invention, with Francis Low, of the `renormalization group' for quantum electrodynamics. With James Hartle he made contributions to quantum mechanics as it would apply to the early universe. He believed that String Theory had the ingredients to be a unifying theory of all interactions including gravity and provided sustained support and encouragement for this activity.

Even though his key contributions lay in the discovery of structure, patterns and laws of nature at sub-atomic scales of space-time, he had a deep appreciation for the complexity of physical, biological and social systems at various scales. In his own words, {\it ``When we human beings experience awe in the face of the splendors of nature, when we show love for one another, and when we care for our more distant relatives---the other organisms with which we share the biosphere---we are exhibiting aspects of the human condition that are no less wonderful for being emergent phenomena''} \cite{Nature Comfortable}. In the last three-and-a-half decades of his life he became an active supporter of, and participant in, the study of complex systems. In 1984 he co-founded the Santa Fe Institute devoted to the study of complex systems. 

In the following we give a brief account of Gell-Mann's scientific contributions. It is beyond the scope of this article to detail all of it;  so we have made a selection and even within that we have detailed some contributions more than others. Rather than follow a chronological order we have discussed his contributions area wise.
\footnote{An excellent biography of Gell-Mann is by George Johnson \cite{Strange Beauty}.}

\section{Elementary Particle Physics}
Murray Gell-Mann was one of the early pioneers of the Standard Model (SM) which is one of the great intellectual achievements of physics, beginning with the discovery of the electron in 1897 by J.J. Thompson and (almost) completed with the discovery of the Higgs boson in 2012 at the Large Hadron Collider in CERN, Geneva. It is a phenomenological theory of all known elementary particles and their weak, strong and electromagnetic interactions. Its building blocks are quarks, leptons and the particles that mediate the interactions between them. The idea of quarks was proposed by Gell-Mann (and independently by Zweig), which we discuss in more detail later. The SM is based on the physical principles of special theory of relativity and quantum mechanics and has been tested to distance scales as small as $10^{-16}$ cms at the Large Hadron Collider (LHC) in CERN, Geneva. It is the culmination of the work of thousands of theorists, experimentalists and engineers from all over the world. 

\subsection{Strangeness quantum number and approximate symmetries of the strong interactions} 
In the early 1950s, high energy particles observed in cosmic ray showers and in experiments at Berkeley and Brookhaven labs presented a puzzle. It appeared that there was fast production of pairs of these particles (``associated production'' of pions, kaons and hyperons) each of which however decayed at a rate much slower than the characteristic rate of production. If they did not decay at all, it would have been like electron-positron pair creation. Here they behave like elementary particles at the time scale involved in their production, yet in fact they are not since they do decay eventually. What exactly is the nature of these particles which seem to have a property unlike anything hitherto known?

The puzzle was solved in 1956 by Gell-Mann (and independently
T. Nakano and K. Nishijima), who introduced a new property (`quantum
number') called `strangeness' associated with strongly interacting
particles that is conserved in strong and electromagnetic
interactions but not in weak interactions. The strange
particles thus behave like stable particles at short time scales (characteristic of electromagnetic and strong interactions) but decay at the relatively
slow time scale of the weak interactions \cite{Strangeness}. This new, approximately conserved symmetry, turned out to provide an invaluable organizing principle in classifying the zoo of particles which were discovered in the 50s and 60s. 

Before dwelling further on how this was done, let us look back at a similar phenomenon in nuclear physics prior to strangeness. A few decades earlier the fact that neutron and proton masses are nearly equal prompted Heisenberg to postulate a symmetry, called isospin, of nuclear (strong) interactions; it was assumed that the small mass difference was  due to a small violation of this symmetry in electromagnetic interactions. In Heisenberg's scheme the proton (electric charge = +1) and neutron (electric charge = 0) are regarded as a doublet of `isospin states' (like spin up and spin down states for the electron) that transform into each other under the isospin group SU(2). Similarly, the three (two charged and one neutral) pions are organized as a triplet of $SU(2)$. All strong interaction processes involving nucleons and pions preserve the isospin symmetry. Small departures from this symmetry, such as due to electromagnetic interactions, are treated very much like symmetry breaking in atomic physics in the presence of a magnetic field along a fixed direction in space (Zeeman effect). In addition to isospin conservation, in a nuclear reaction the number of nucleons (protons or neutrons) is also conserved and this book-keeping is done by introducing the conserved baryon number $B$, with the proton and neutron both assigned $B=1$, their antiparticles $B=-1$ and the mesons assigned $B=0$. Once the isospin and baryon number of a particle are known, its electric charge is given by the formula: $Q = I_3 + B/2$, where $I_3$ is the 3rd component of isospin. This gives the electric charge of the proton as $1/2 + 1/2= +1$ and that of the neutron as $-1/2 + 1/2=0$, as required.

How does strangeness fit into this scheme of things? Nakano, Nishijima
and Gell-Mann proposed that the formula for the electric charge must
become 
\begin{equation}\label{NNG}
Q = I_3 + (B + S)/2 
\end{equation}
(known as the Gell-Mann-Nakano-Nishijima formula), where $S$
is the strangeness quantum number. The quantity $Y=B+S$ is given the
name hypercharge. When particles of the same spin, parity and baryon
number were assigned isospin and strangeness from observed decay
reactions and plotted in the $(I_3 , Y)$-plane, they revealed
recognizable geometric patterns for both mesons and baryons (see
figure), in a way akin to the periodic table of chemical
elements. These patterns were later recognized as diagrams related to
the symmetry group $SU(3)$.

\begin{figure}
\begin{center}
\fbox{\includegraphics[width=0.6\linewidth]{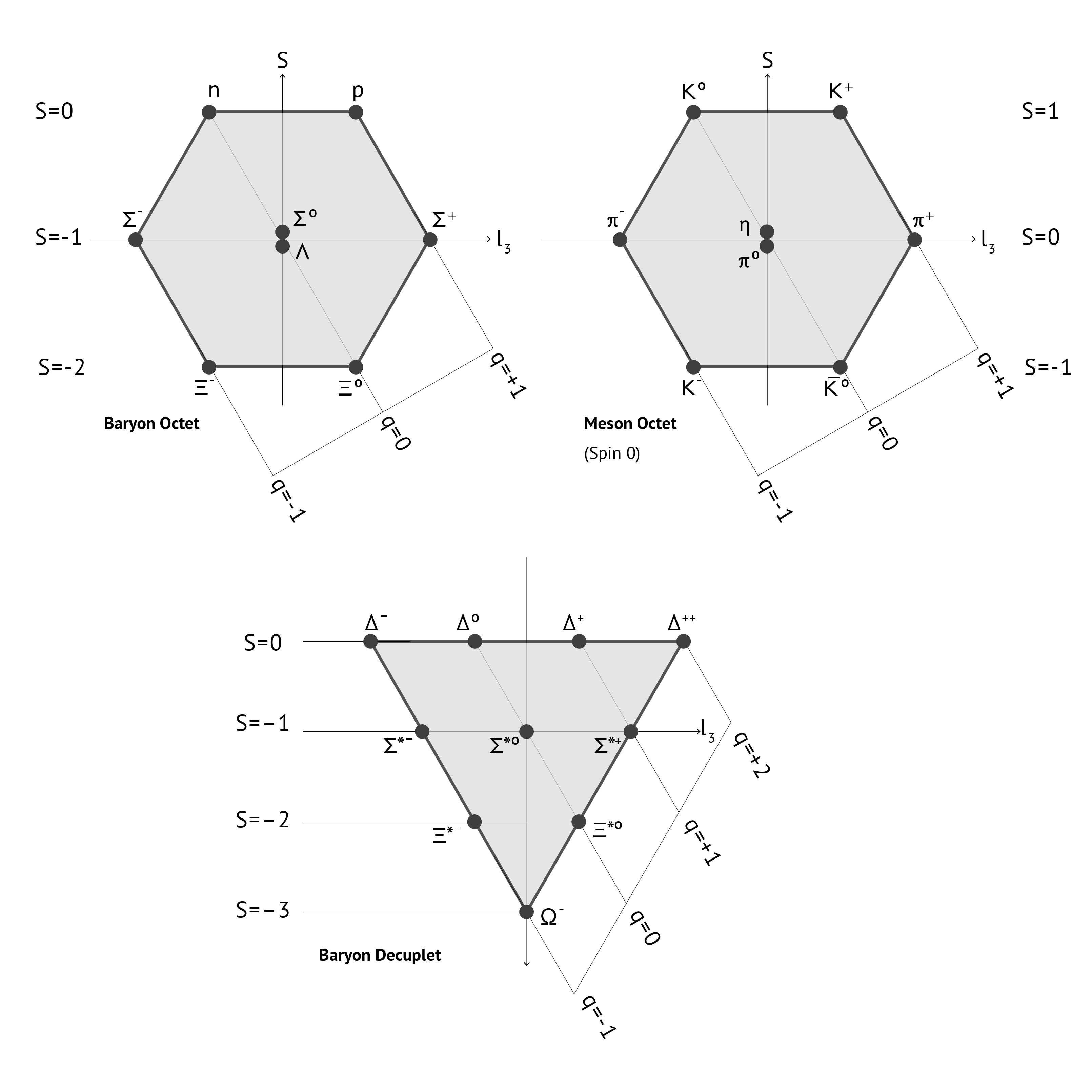}}
\caption*{\it Eight Fold Way description of the baryon and meson octet and the baryon decuplet.\\ (Figure credit: Juny Wilfred, ICTS)}
\end{center}
\end{figure}

 \subsection{The Eight-fold way}
It took a while and some learning of mathematics to recognize that the Gell-Mann-Nakano-Nishijima formula for electric charge is a sum of the two commuting generators of the Lie group SU(3) which corresponded to isospin and hypercharge. This fundamental discovery was made independently by Gell-Mann and Yuval Ne'eman in 1961 \cite{Gell-Mann Ne'eman}. The pseudo-scalar and vector mesons are organized by the octet representation (8 states), and the baryons by the octet (8 states) and decuplet (10 states) representations of SU(3). This classification was done assuming that SU(3) is an exact symmetry, neglecting small violations by the weak and electromagnetic interactions.

The approximate SU(3) symmetry was powerful enough to lead Gell-Mann (1961) and independently S. Okubo (1962) to arrive at their celebrated mass formula \cite{Gell-Mann-Okubo} for baryons:
\begin{equation}\label{eq:Liouville-startingGell-Mann-Okubo} 
M = (a_1 + a_2 Y +a_3 (I(I+1) - Y^{2}/4))
\end{equation}
(known as the Gell-Mann-Okubo formula), where $Y$ is the hypercharge and $I$ is the total isospin of a given representation of SU(3). The parameters $a_i$ $i=1,2,3$ depend on the representation and are experimentally determined. This formula applied to baryons works within $0.5 \%$ of the measured values of the baryons it describes! For mesons there is a similar formula but for $M^2$. The reason for the agreement of the Gell-Mann-Okubo (G-O) formula with known particle data was not understood until decades later. The G-O formula made an important prediction that  the baryon decuplet should contain the spin 3/2 strange baryon $\Omega_{-}$ with $S = -3$ and mass $1672$ MeV. Sure enough, the particle was experimentally found at the same mass by the group led by Nicholas Samios at Brookhaven in 1964 \cite {Omega-}. This was a great triumph of the eight-fold way. The organization in terms of the SU(3) Lie group, with  eight generators, was named the `eight-fold way'. The name, due to Gell-Mann, is taken from the `Eight-Fold Path' to enlightenment (Astangika-Marga in Sanskrit) said to be announced by the Buddha in his first sermon.

\subsection{The Quark Model}
The great success of the eight-fold way based on the approximate symmetry group SU(3) led to a prescient observation by Gell-Mann and Zweig in 1964 \cite{Quark}. They noted a fact from group theory that the meson and baryon state quantum numbers can be explained very simply in terms of the more elementary (fundamental and anti-fundamental) representations of SU(3) involving three states and their conjugates in terms of which all other representations can be built. Gell-Mann and G. Zweig independently made the bold proposal that these states may correspond to elementary spin 1/2 fermions called `quarks' (by Gell-Mann) and `aces' (by Zweig). Gell-Mann's nomenclature, which is now standard, was inspired by his reading of `Finnegans Wake', by James Joyce, where he came across the word `quark' in the phrase ``Three quarks for Muster Mark''.  

It should be pointed out that the idea that all elementary particles should be composed of some which are more elementary was originally considered in 1955 by Shoichi Sakata. The Sakata Model considered the lightest baryons $(p, n, \Lambda)$ as fundamental and transforming in the fundamental representation of SU(3). The anti-particle triplet transforms under the anti-fundamental representation. Unlike the quark model the Sakata Model could not account very naturally for the baryon octet and decuplet. For a review see Lev Okun's article \cite{okun}.

The electric charges of the quarks are fixed by the NNG formula to be fractional, namely $2/3$ or $-1/3$ in units of the charge of the electron, with opposite charges for anti-quarks. This was contrary to the accepted wisdom because fractional electric charges had never been observed, and besides, the single-valued-ness of the wave function of an {\it isolated charge} requires the electric charge to be an integral multiple of the charge of an electron. 

As no one had experimentally seen quarks, they were treated as a mathematical device to account for the quantum numbers of the observed hadrons. The 3-species of quarks were labeled up (u), down (d) and strange (s). The u and d quarks have strangeness equal to zero, while the s quark carries strangeness $S=-1$. For example, the proton would be built of (uud), the neutron (udd) and $\Lambda$ (uds). The strange baryon $\Omega_{-}$ would be (sss) and so on. The breaking of the approximate SU(3) symmetry can now be ascribed to the strange quark being more massive than the u and d quarks. 

There were two puzzling aspects of this proposal. The first was that the constituent description of $\Omega_{-}$ led to an apparent contradiction with the Pauli exclusion principle for fermions since three identical fermions are in the same state. The second concerns the fact mentioned above: that quarks were never seen experimentally. 

\subsection{The Color of Quarks and Quantum Chromodynamics (QCD)}
From the quark model to asymptotic freedom and quark confinement is a heroic tale of scientific discovery. There were strongly held views in the 1960s about the inadequacy of quantum field theory to describe the strong interactions and even stronger views that quantum field theory may not exist as a consistent mathematical theory. Alternative formalisms like S-Matrix theory, dispersion relations, current algebra and effective field theory were pursued to do meaningful calculations. Gell-Mann contributed to all these topics including the proposal of current algebra in 1964. 

Continuing our narrative, there were two proposals to resolve the  $\Omega_{-}$  puzzle within the quark model. W. Greenberg proposed that quarks obey an order three para-statistics. However, Y. Nambu's simpler and far reaching proposal was that each of the (u,d,s) quarks came in additional colors. Han and Nambu introduced the color triplet model in which each quark carried an additional quantum number, and the quarks transform under a new symmetry group (`color' ) $SU(3)_c$ \cite{Han:1965pf}. The fermions in $\Omega_{-}$  can be in the same energy state but distinguished by their colors consistent with the Pauli principle. Now that quarks have color, the original labels u, d and s are called  quark `flavors'. Hence quarks have two labels `flavor' and `color'. 

The solution to the second puzzle that `free quarks' had not been observed, was an intense effort in both experiment and theory for more than a decade. Nambu proposed that the forces between colored quarks could be mediated by the $SU(3)_c$ octet of eight `gluons', similar to the photon mediating the force between charged particles in quantum electrodynamics \cite{Nambu Color1}. 
However, he did not have a clear formulation of the corresponding non-Abelian gauge theory \cite{Nambu Color2}. 
Subsequently Fritzsch and Gell-Mann also suggested a $SU(3)_c$ gauge theory (among other ideas including singlet gluons) \cite{Gell-Mann Fritzsch}. Gell-Mann seems to have been reluctant to accept quarks as a real particles except as a crutch to extract a correct theory of the strong interactions. Wrapping up his talk at the 
XVI International Conference on High Energy Physics in Chicago in 1972, he said {\it ``Let us end by emphasizing our main point, that it may well be possible to construct an explicit theory of hadrons, based on quarks and some kind of glue, treated as fictitious, but with enough physical properties abstracted and applied to real hadrons to constitute a complete theory"} \cite{Gell-Mann Fritzsch}.

What was missing was an understanding and a formulation of an appropriate question and calculation of the dynamics of interacting quarks and gluons. In the 1960s and the early 1970s no one could be sure among the various proposals, which one was the correct theory of the strong interactions. There was no evidence for quarks, and gluons were on the face of it massless and unseen. Besides, the existence of quantum field theory was in disrepute due to severe criticisms of its existence especially from the Landau school based on the fact that in the theory of electrons interacting via the electromagnetic field, Quantum Electrodynamics (QED), the effective electric charge grows unbounded at short distances or equivalently high energies. 

Guidance now came from the results of the MIT-SLAC experiments led by Friedman, Kendall and Taylor at the Stanford Linear Accelerator Center in 1969 involving high energy inelastic scattering of electrons off protons and neutrons \cite{MIT-SLAC}.
The data exhibited Rutherford type wide angle scattering of the electrons indicating the presence of point like structures inside protons and a scaling law at high energies, which could be explained by R. P. Feynman's hypothesis that hadrons consisted of point-like weakly interacting structures called partons. The scaling law was arrived at earlier using more formal methods by J. D. Bjorken. The point like partons were eventually identified with quarks after the discovery of asymptotic freedom as we discuss below. These experiments and K. G. Wilson's ideas on the renormalization group inspired a search for a quantum field theory that would be able to explain the observed scaling \cite{Wilson}. 

What theory could account for the strange combination of facts --- nucleons when probed at high energies seem to behave as if they are constituted by weakly interacting point-like particles, which at low energies are so strongly bound that they cannot even be observed separately? It is the discovery of asymptotic freedom in 1973 by Gross and Wilczek, and Politzer in non-abelian gauge theories, that firmly established QCD as a theory of the strong interactions. Asymptotically free theories exist and exhibit a logarithmically vanishing coupling constant $g(E) \sim 1/\log{(E/\mu)}$ for energies $E \gg \mu$, where $\mu$ is a characteristic scale of the theory. Asymptotic freedom restored confidence in quantum field theory. These theories could explain the experiments and also gave rise to the expectation that as energies decrease the coupling constant grows, indicating the possibility that quarks are permanently confined within the hadrons \cite{Gross Wilczek}. This was subsequently demonstrated in 1974 by K. G. Wilson in the strong coupling expansion of a non-perturbative formulation of the non-abelian gauge theory on a euclidean space-time lattice. Subsequent work that followed has established a cross-over from asymptotic freedom to quark confinement, solving one of the great mysteries of the strong interactions. 

In conclusion when the dust settled the theory of the quarks of Gell-Mann and Zweig with fractional electric charge interacting via an octet of gauge fields (gluons) of $SU(3)_c$ color, came to be established as the correct theory of the strong interactions in which the partons could be identified as quarks which always remain confined in the hadrons. This theory was called `quantum chromodynamics' (QCD) by Gell-Mann. As time went on, heavier quarks were discovered experimentally. They are named charmed (c), top (t) and bottom (b) and the standard model today is organized in terms of three doublets (generations) of quarks: (u,d), (c,s) and (t,b) of increasing mass. These together with the leptons (electron, muon, tau and their associated neutrinos constitute the known matter part of the Standard Model.

George Johnson summarizes Gell-Mann's feelings about QCD in the following way \cite{Strange Beauty}:
{\it ``After all the exotic avenues physicists had explored in search of a theory of the strong force  --- dispersion relations, the bootstrap, string theory --- the shape of the answer that emerged was essentially old-fashioned QED, with a twist called asymptotic freedom. The result, far more elegant than the lopsided electroweak theory, was a theoretical masterpiece that Gell-Mann would have loved to call his own. Quarks were his and Zweig's, the Eightfold Way was his and Ne'eman's. He could even content himself with knowing that buried within asymptotic freedom was his and Low's old work on the renormalization group. But he was filled with regret that he hadn't seen the whole thing. The new picture of the strong force was the crowning vindication of both quarks and field theory, ideas Murray had been so deeply ambivalent about."}

\subsection{The V-A weak interaction}
Gell-Mann worked on the V-A  structure of the weak interactions with Richard Feynman (1958). E. C. George Sudarshan and Robert Marshak had independently and earlier discovered, based on a careful analysis of experimental data, the V-A structure that incorporates maximal parity violation. This seminal discovery paved the way for the work on gauge unification of electro-weak interactions by Sheldon Glashow, Abdus Salam and Steven Weinberg. We do not detail this discovery as it has been discussed in a recent article co-authored by one of us on Richard Feynman \cite{V-A}. Other excellent accounts of the V-A story are by S. Weinberg \cite{Weinberg} and by George Johnson \cite{Strange Beauty}. 

\subsection{The Renormalization Group}
The renormalization group is without doubt one of the centre-pieces of quantum field theory (QFT). It was developed by Gell-Mann and Francis Low (and independently a bit earlier by Stuekelberg and Petermann) 
\cite{Gell-Mann Low}.
It goes to the heart of the issue of the definition of a QFT (here QED), which is defined with the introduction of an ultraviolet momentum cut-off $\Lambda$ so that one considers only distances $d \ll \frac{1}{\Lambda}$ (in natural units where $\hbar = c =1$).
In such a case the calculated answers for various physical quantities like the potential between the two charged electrons are finite. But questions remain about the physical significance of the (arbitrary) cutoff scale that is introduced to define the theory. Gell-Mann and Low introduced the notion that the electric charge of an electron $e$ depends on the distance $r$
at which one observes it, and they showed that the change in the value of the charge as a function of the distance or equivalently the momentum scale $\mu = \frac{1}{r}$
satisfies a first order `renormalization group' differential equation 
\begin{equation}\label{eq:Liouville-startingGell-Mann-Low}
\mu\frac{d e(\mu)}{d\mu} = \beta (e(\mu)).
\end{equation}
$\mu$ here is in units of $m_e$ the mass of the electron. \footnote {Originally the Gell-Mann-Low equation was for $e{^2}(\mu)$ and their $\psi$ function is given by $\psi (x) =x\beta(x)$.}

The function $\beta$ can be calculated in a power series in $e{^2}(\mu)$, with coefficients that are independent of $\mu$. The leading order result of the beta function is 
\begin{equation}\label{QED beta}
\beta(e(\mu)) = \frac{1}{12\pi^2}e{^3}(\mu).
\end{equation}
Integrating the Gell-Mann-Low equation one can relate the electric charge at two different scales $\mu$ and $\mu_0,$
\begin{equation}\label{QED charge}
e{^2}(\mu) =  \frac{e{^2}(\mu_0)}{(1 - \frac{e{^2}(\mu_0)}{6\pi^2}log(\frac{\mu}{\mu_0}))}.
\end{equation}
Due to the presence of the minus sign in the denominator, which reflects the fact that the beta function is positive, the electron charge $e(\mu)$ increases as $\mu$ approaches $m_e$ from below (or equivalently the distance scale $r$ approaches the Compton wavelength of the electron) and QED becomes strongly coupled. 

Gell-Mann and Low also made the important observation that the vanishing of the $\beta$ function would imply that the renormalization group equation has a {\it fixed point} where the electric charge would be a constant. 

The seminal work of Gell-Mann and Low in QED implied that quantum field theory becomes strongly coupled at short distances and hence may not exist, a viewpoint advocated by the Landau school. It took many years before the renormalization group and quantum field theory attained their central position in methods to explore systems with many degrees of freedom. Notable here is K. G. Wilson's work that `explained' both quantum field theory and critical phenomena in condensed matter systems \cite{Wilson} and the discovery of asymptotic freedom \cite{Gross Wilczek}. The property of asymptotic freedom mentioned earlier is precisely the fact that in QCD the perturbative beta function is negative and hence the coupling constant goes to zero at high energies. QED, inconsistent on its own, is rendered consistent in the Standard Model by virtue of being embedded in a theory with a non-abelian gauge group.
\newpage

\section{Unification of all interactions and String Theory}
Gell-Mann believed that a truly unified theory must contain the force of gravity besides the weak, strong and electromagnetic force. He was persuaded that String Theory had the ingredients of a unified theory of all interactions. The basic ingredients of the supersymmetric version of String Theory (Superstring Theory) were discovered around 1971 (by P. Ramond, J. Schwarz and A. Neveu, and J. L. Gervais and B. Sakita). In these early years as String Theory was taking roots he created at Caltech what he called ``a nature reserve for an endangered species'' \cite{Strange Beauty}. This reserve supported with enthusiasm for a long time the careers of Lars Brink, Michael Green, Pierre Ramond and John Schwarz who made important contributions to the subject. 
This period also witnessed the discovery in 1974 (by T. Yoneya, and J. Scherk and J. Schwarz) that String Theory contains Einstein's gravity, and the discovery in 1984 (by M. Green and J. Schwarz) of anomaly cancellation in Superstring Theory, which makes it a candidate for a unified theory of all interactions including gravity.

\section{Quantum Mechanics and the Universe}
Gell-Mann was worried about ``how to fit history into quantum 
description'' already in 1962 
\footnote{Recalls Virendra Singh, who spent time at Caltech with Gell-Mann.}. Many years later, beginning in the early 1990s, Gell-Mann and Hartle, (see e.g. \cite{Gell-Mann:2013hza}, \cite{Hartle Solvay} and references therein) undertook a systematic exploration of the quantum description of the universe, which is a `closed quantum system' by definition. In such a framework, the world we observe, the instruments of observation and the environment are all part of one quantum system. The usual `Copenhagen' rules draw a sharp distinction between the observed quantum system and the `observing apparatus', which is required to obey the rules of classical mechanics. This is undoubtedly unsatisfactory as it not only alludes to an incompleteness of quantum mechanics but also makes it inapplicable to the early universe when there was no one observing the universe.   

Using understanding gained in earlier studies of closed quantum systems, they represented the full wave function for a quantum process, starting from a given initial state, as a sum over classes of `coarse grained' histories. These histories are built out of a mutually exclusive and exhaustive set of Feynman paths, which themselves are called fine-grained histories.  Let us denote by $|\Psi_{\alpha}\rangle$ the ``branch state vector'' which is the sum over Feynman paths belonging to the class $\alpha$. A description of quantum mechanics in terms of coarse grained histories is called
``generalized quantum mechanics''. The full wave function, which is a sum over {\it all} Feynman paths is naturally the sum over all branch state vectors: $|\Psi\rangle = \sum_\alpha |\Psi_{\alpha}\rangle$. Then 
$\langle \Psi |\Psi\rangle$ ($= 1$) consists of a sum over the branch probabilities $p_\alpha = \langle \Psi_\alpha |\Psi_\alpha\rangle$ plus the quantum mechanical interference terms $ D(\alpha', \alpha) = \langle\Psi_{\alpha'}|\Psi_{\alpha}\rangle$ where $D$ is referred to as the decoherence functional. If the decoherence condition, $ D(\alpha', \alpha) = 0$ for all $\alpha \not= \alpha'$, is satisfied, the coarse grained histories are said to decohere, and we recover the classical probability sum rule $1=\sum_\alpha p_\alpha$. This viewpoint then allows one to recover the Born rule without needing to introduce non-unitary measurement processes. The question then arises as to whether and when the decoherence condition is satisfied. It appears that in a large class of closed systems, there are natural choices of the coarse grained histories such that the decoherence condition is indeed satisfied to a very good approximation due to interactions between parts of the system. This suggests that the framework of `generalized quantum mechanics'  obviates the need to introduce any external observing apparatus and allows one to understand and interpret quantum mechanics of the universe as a whole.

\section{Complex Systems}
As a child and during his school-going years, Murray Gell-Mann was not particularly interested in physics. In his own words \cite{Caltech oral archive}, {\it ``My principal interests were all in subjects involving individuality, diversity, evolution. History, archeology, linguistics, natural history of various kind --- birds, butterflies, trees, herbaceous flowering plants, and so on --- those are the things that I loved. Plus mathematics $\ldots$ I tried reading physics books. I had great difficulty with them $\ldots$ the different subjects were not connected --- mechanics, wave motion, sound, light, heat, electricity, magnetism. One would never have known that there were a few simple laws that governed all of these things."} Gell-Mann described the choice of physics as his undergraduate subject at Yale as a compromise between his own preferences --- archaeology or linguistics --- and his father's --- engineering. 

In later years while he was making his seminal contributions to fundamental physics, his childhood interests never left him. He amassed an encyclopaedic knowledge of world cultures, religions, languages and history. The last three and a half decades of his life marked a more active return to those interests through his involvement with the Santa Fe Institute and science of complex systems.

The Santa Fe Institute (SFI) was founded in 1984 by a group of people led by George Cowan of the Los Alamos National Laboratory, ``to bring the tools of physics, computation and biology to bear on the social sciences", breaking disciplinary boundaries to ``seek insights that were useful for both science and society" \cite{Germain}. Gell-Mann was a member of the founding collective and he attracted a large number of outstanding scholars in diverse disciplines to the Institute. Physicist and Nobel laureate Philip W. Anderson said this of SFI in 1992 \cite{PWA}, {\it ``From the first we had probably the highest ratio of scientific eminence, commitment and sheer competence to physical plant and actual funding since Galileo's Accademia dei Lincei, or Academy of the Lynx-Eyed (that is, the far-seeing)."} This enabled SFI to spearhead the science of complex systems and establish it as a central concern of our times. In addition to attracting eminent people and encouraging younger colleagues, Gell-Mann also carried out research in complex systems at SFI. 

Gell-Mann was a firm believer in science and reason. Till the end, he was in pursuit of a scientific understanding of various aspects of nature, from the fundamental laws and initial conditions of the universe to the complex world we see around us. He believed that calling a fundamental law the `theory of everything' is inappropriate. `Everything' according to him is a consequence of not just fundamental law and the initial condition of the universe, but, in addition, chance events. He was very clear that knowing the fundamental laws at sub-atomic scales does not mean that we can explain 
``the diversity of nature and the striking way that diversity is organized". There is a hierarchy of laws and conditions that is needed to explain the natural world and its multi-dimensional evolution from elementary particles, to atoms and molecules, to living things, to conscious beings such as us. He has made a great effort to elaborate on this view of the natural world in his book, `The Quark and the Jaguar' \cite{Quark Jaguar}. For some examples and properties of complex systems, see box.\\

\fbox{
\parbox{15.0truecm}{{\bf What are complex systems?} There is no satisfactory crisp definition of a complex system, but significant examples include living organisms, ecosystems, the brain, and human societies. These systems are made up of many components. In the above examples, we might take the components to be, respectively, molecules, biological species, neurons, and human beings. Unlike typical many-body systems studied in physics, the components are non-identical. The interactions of the components with each other can be represented by a network that is neither regular nor completely random, but structured. This endows each component with a specific `role' in the short-term dynamics of the system. In the long term, the system `evolves' as existing components are replaced by others, the network changes, new roles are created and qualitatively new system-level properties emerge. In these systems the set of degrees of freedom (or time dependent variables) is itself a variable. The systems are constantly exchanging matter and energy with their environment, and as they evolve, they typically also transform their environment. The systems are typically robust to fluctuations, but sometimes, even small fluctuations can cause them to collapse. \\ \\
These systems are increasingly being probed using a combination of quantitative methods from physics, mathematics and engineering. Major questions about complex systems include the following: How does one understand the behaviour of the whole from a knowledge of its parts? How does one identify the salient degrees of freedom and the regularities displayed by the system at different length and time scales, or equivalently, how does one coarse-grain these systems in a useful manner? How does one construct structural and dynamical models for these systems that are useful for quantitative analysis, measurement and prediction? How did these systems originate? How do they grow in complexity? Why do they crash? How does one quantify `innovation' and its consequences in these systems?}\\
}

\subsection{Measures of Complexity}
It is difficult to precisely characterize the `complexity' of these systems. This is one of the questions that interested Gell-Mann. When can one call an entity complex? Consider for concreteness the set of binary strings. A notion of complexity of a string, called `algorithmic complexity', is the length of the smallest computer program or algorithm that will generate that string. A long but ordered string of length $L$ such as $1111\ldots 1$  has low algorithmic complexity because a very small program  can generate it (e.g., PRINT $L$ ONES, whose length, $\sim \log L$, is much smaller than $L$). However, a random string, such as one describing a long sequence of coin tosses, has large algorithmic complexity proportional to its length (since each successive entry can only be generated after the result of the coin toss is known, the random sequence is its own minimal program). One would like a complexity measure that assigns a low complexity to such random strings as well. Using ideas from information theory, statistical physics and theoretical computer science, Gell-Mann and Lloyd \cite{SethLloyd,Complexity1} proposed two measures, `effective complexity' ${\cal{E}}$ and `total information' $\Sigma$. Given a putative model (in general with both deterministic and stochastic features) for generating the entity in question, or, equivalently, an ensemble of entities similar to the entity in question along with the probability for each member of the ensemble, the `effective complexity' of the entity is defined as the length of the shortest description of the model or ensemble. `Total information' of the entity is the sum of its effective complexity and the entropy of the ensemble. While both quantities depend upon the `eye of the beholder' (in that they depend upon the model or ensemble chosen to describe the entity), Gell-Mann and Lloyd argued that the model that minimizes $\Sigma$, and subject to that constraint, also minimizes ${\cal{E}}$ is a particularly interesting one. For such a model, effective complexity is approximately equal to the amount of information required to describe the entity's regularities and corresponds to our natural intuition about complexity: entities generated by easily specified processes, whether deterministic or wholly random (such as the two strings mentioned above), are simple, while those that require a combination of deterministic processes and a large number of historical accidents are complex. 

Gell-Mann also worked on other aspects of complex systems at SFI. This includes work on `non-extensive entropy' \cite{Tsallis}, generalized entropies \cite{Hanel}, and gambles in economics \cite{Peters}. He also played an active role in supporting the program on human languages at SFI. 

As a fitting tribute to a great scientist and thinker of our times, the Santa Fe Institute's main building is named after Murray Gell-Mann. 

\begin{figure}
\begin{center}
\fbox{\includegraphics[width=0.6\linewidth]{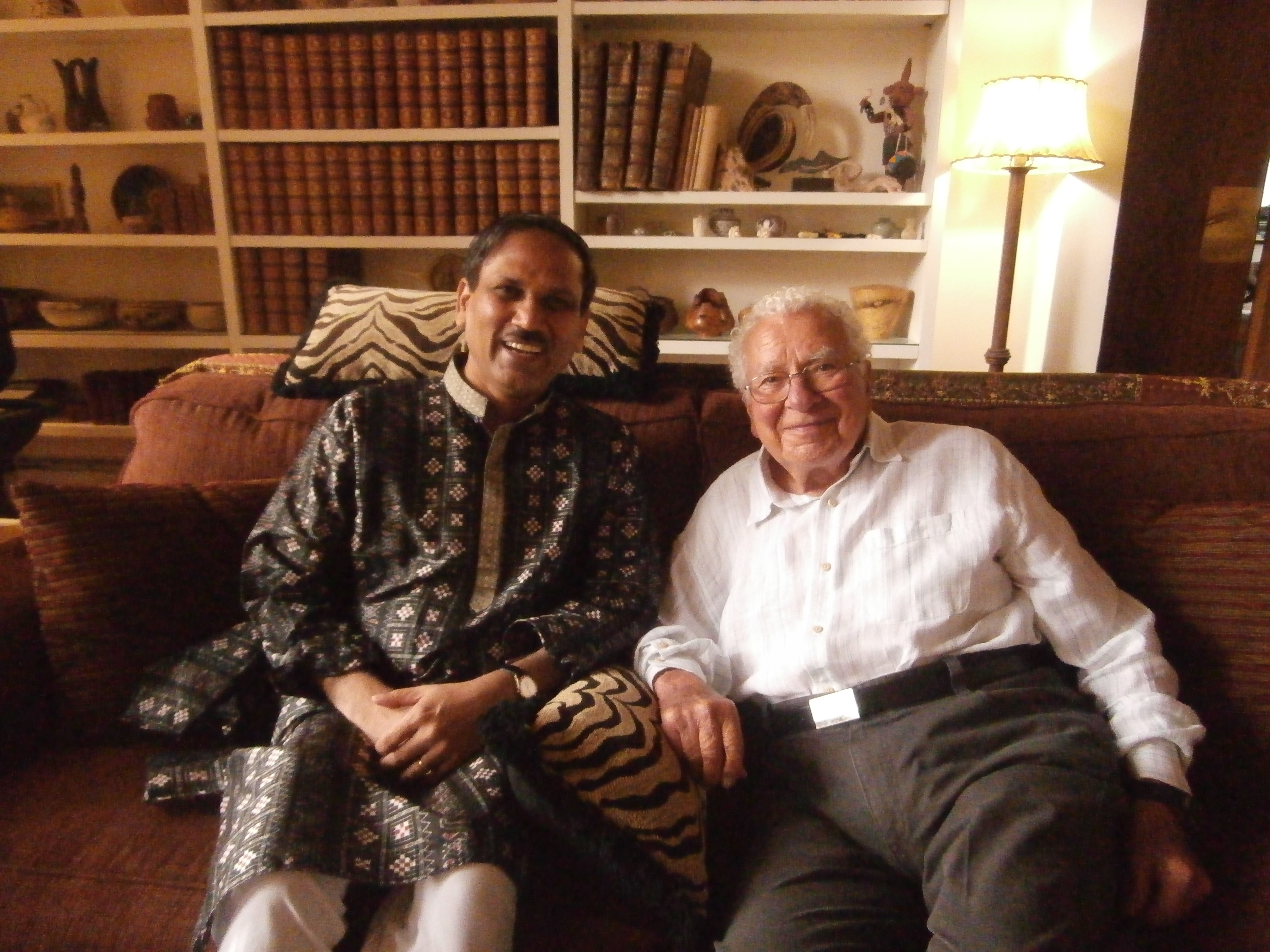}}
\caption*{\it Murray Gell-Mann at his home in Santa Fe, 2014, with one of the authors (SJ).}
\end{center}
\end{figure}

\section{Gell-Mann's visit to India}

\begin{figure}
\begin{center}
\fbox{\includegraphics[width=0.6\linewidth]{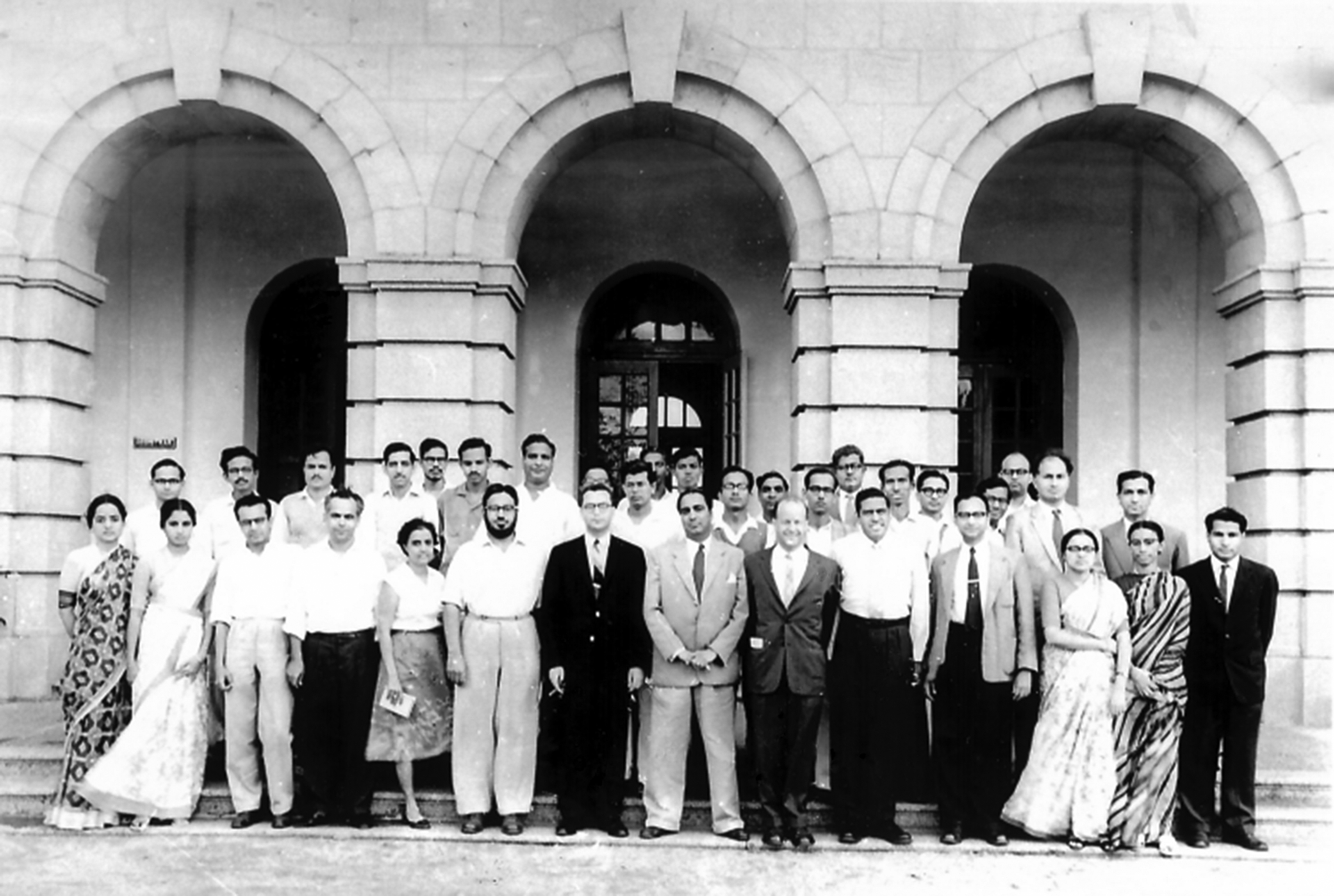}}
\caption*{\it TIFR Summer School at IISc Bangalore, 1961. Murray Gell-Mann is flanked on his left by Homi J. Bhabha, Founder of the Tata Institute of Fundamental Research. (Photo Credit: TIFR Archives).}
\end{center}
\end{figure}

Gell-Mann had visited India and lectured on `Weak Interactions of Strongly Interacting Particles' at the 
Summer School on Theoretical Physics at IISc Bangalore, organized by the Tata Institute of Fundamental Research (TIFR) in 1961 \cite{TIFR summer school notes}. He subsequently co-authored papers on Regge Pole theory with B. M. Udgaonkar and Virendra Singh of TIFR.
\bigskip

\noindent Gell-Mann was elected a Fellow of the Indian National Science Academy, New Delhi, in 1984.

\subsection*{Acknowledgments}
We would like to thank Avinash Dhar, Rajesh Gopakumar, Hirosi Ooguri and Virendra Singh for very useful comments on the article, and Gautam Mandal for many discussions and inputs, in particular on the section dealing with `Quantum Mechanics and the Universe'. 
SRW would like to acknowledge the Infosys Foundation Homi Bhabha Chair at ICTS-TIFR and the hospitality of the Bellagio Center of the Rockefeller Foundation in Como, Italy.

\end{document}